\documentclass{Interspeech}

\interspeechcameraready

\title{Enhancing GOP in CTC-Based Mispronunciation Detection with Phonological Knowledge}

\author{Aditya Kamlesh}{Parikh}
\author{Cristian}{Tejedor-Garcia}
\author{Catia}{Cucchiarini}
\author{Helmer}{Strik}

\affiliation[nocounter]{Centre for Language Studies}{Radboud University}{the Netherlands}
\email{aditya.parikh@ru.nl, cristian.tejedorgarcia@ru.nl, catia.cucchiarini@ru.nl, helmer.strik@ru.nl}

\usepackage{xcolor}

\usepackage{tabularx}
\usepackage{booktabs}
\usepackage{multirow}

\usepackage{comment}
\usepackage[utf8]{inputenc} 
\usepackage{graphicx} 
\usepackage{amsmath} 
\usepackage{amssymb} 
\usepackage{array} 
\usepackage{enumitem} 
\usepackage{tipa} 
\usepackage{booktabs}

\begin{document}

\maketitle

\begin{abstract}

Computer-Assisted Pronunciation Training (CAPT) systems employ automatic measures of pronunciation quality, such as the goodness of pronunciation (GOP) metric. GOP relies on forced alignments, which are prone to labeling and segmentation errors due to acoustic variability. While alignment-free methods address these challenges, they are computationally expensive and scale poorly with phoneme sequence length and inventory size. To enhance efficiency, we introduce a substitution-aware alignment-free GOP that restricts phoneme substitutions based on phoneme clusters and common learner errors. We evaluated our GOP on two L2 English speech datasets, one with child speech, My Pronunciation Coach (MPC), and SpeechOcean762, which includes child and adult speech. We compared RPS (restricted phoneme substitutions) and UPS (unrestricted phoneme substitutions) setups within alignment-free methods, which outperformed the baseline. We discuss our results and outline avenues for future research.
    
\end{abstract}

\section{Introduction}

Language is a fundamental skill that shapes communication, cognitive development, and cultural integration \cite{nardon2012language} and learning languages other than the native one is essential in our globalized society. Traditional classroom settings often make it difficult for teachers to provide the degree of individualized attention that is required for high-quality language learning, especially when it comes to speaking and pronunciation \cite{onyishi2020teachers, neri2002feedback}. However, early and effective pronunciation training can help language learners improve their ability to master the language \cite{cucchiarini2009oral}.

CAPT systems that incorporate Automatic Speech Recognition (ASR) technology can offer promising solutions to these challenges by providing personalized and scalable learning experiences \cite{cucchiarini2009oral, neri2006asr}. These systems enable learners to practice independently through "read-aloud" exercises, incorporating Mispronunciation Detection and/or Diagnosis (MDD) to deliver corrective feedback at the phoneme, word, and sentence levels \cite{Amrate_Tsai_2025, liu2016teaching}. Such targeted feedback can help learners bridge the gap between their L1 and L2 while promoting accurate pronunciation and confident communication \cite{silpachai2024corrective, doremalen2013automatic}.

The goodness of pronunciation (GOP) is a widely used measure in CAPT research that quantifies pronunciation quality by analyzing posterior probabilities from an ASR system's acoustic model  \cite{witt2000phone, witt2000use}. Initially developed using Hidden Markov Models (HMMs), GOP has since evolved based on deep neural networks (DNNs). Variants such as weighted-GOP \cite{van2010using}, lattice-based GOP \cite{song2010lattice}, and context-aware GOP \cite{shi2020context} have further improved accuracy and robustness.
More recently, the emergence of Self-Supervised Learning (SSL) models, such as Wav2vec2.0, Hubert, and WavLM, has significantly advanced MDD and phoneme recognition tasks \cite{baevski2020wav2vec, hsu2021hubert, chen2022wavlm}. These pretrained models extract rich speech embeddings and require less labeled data for fine-tuning compared to the traditional supervised approaches. In MDD, they are used to compare canonical transcriptions with phoneme recognition outputs to identify pronunciation errors. However, this process involves challenges related to sequence alignment and the need for explicitly labeled mispronunciation data, which is expensive to produce. 

In the Connectionist Temporal Classification (CTC) framework, GOP scores for SSL fine-tuned phoneme recognition models can be computed using either forced alignment or alignment-free approaches \cite{medin2024self, cao2024framework}. Forced alignment can be unreliable in non-native speech due to acoustic variability, potentially overlooking pronunciation errors \cite{mathad2021impact}. The alignment-free method by Cao et al. \cite{cao2024framework} computes GOP features using scalar and multi-dimensional vector representations, improving the handling of substitution and deletion errors. However, in multilingual phoneme recognition with large phoneme inventories, this becomes computationally expensive, since phonemes in the canonical transcript must be substituted or inserted with others in the phoneme inventory, exponentially increasing computational costs, memory demands, and false positives. As the phoneme inventory expands, the CTC graph complexity grows quadratically \cite{laptev2021ctc}, making real-time pronunciation learning impractical since immediate feedback is required.

To address these challenges in alignment-free CTC-based MDD for large phoneme inventories, we propose incorporating phoneme clustering \cite{tak2010clustering,meng2009multilingual,oh2021hierarchical} and learner-specific error modeling \cite{cucchiarini2011error,kruitbosch2020pronunciation}. By grouping acoustically and articulatory similar phonemes, phoneme clustering reduces computational complexity while preserving essential phonemic distinctions. Integrating knowledge of common substitution and deletion errors made by non-native speakers can further refine MDD.

To the best of our knowledge, no prior work has applied phoneme clustering and learner-specific error modeling in an alignment-free CTC-based MDD system. To investigate the potential of our novel approach, we conducted a study that addressed the following research question (RQ): How do phoneme clustering and learner-specific error modeling affect the performance of an alignment-free CTC-based MDD system compared to unrestricted phoneme substitutions?

\section{Methodology}

\subsection{GOP Definitions}
First, we follow the definition of GOP by Witt and Young \cite{witt2000use}. They compute GOP using the sequence of feature vectors \( \mathbf{O}_1^T = \{\mathbf{o}_1, \ldots, \mathbf{o}_T\} \) of length \( T \) and the corresponding canonical phoneme transcription \( L_{\text{cano}} = \{l_1, \ldots, l_{|L_{\text{cano}}|}\} \). For a given phoneme \( l_i \in L_{\text{cano}} \), the original definition of GOP is based on the log-posterior probability of that phoneme:

\begin{equation}
\text{GOP}_{\text{classical}}(l_i) = \log \left( \frac{p(\mathbf{O}_{t_1}^{t_2} \mid l_i) P(l_i)}{\sum_{q \in \mathcal{Q}} p(\mathbf{O}_{t_1}^{t_2} \mid q ) P(q)} \right) \bigg/ (t_2 - t_1),
\label{eq:gop_classical}
\end{equation}

where \( \mathbf{O}_{t_1}^{t_2} \) represents the feature frames corresponding to the canonical phoneme \( l_i \), and \( \mathcal{Q} \) is the set of all phonemes in the target language. The term \( p(\mathbf{O}_{t_1}^{t_2} \mid l_i)\) is typically modeled by HMM-GMM-based acoustic models. The score is then normalized by the duration of the segment, \( t_2 - t_1 \). This method relies on forced alignment using ASR to align the canonical transcription to the speech recording. The resulting GOP scores indicate deviations in pronunciation for each phoneme.

Second, with DNNs, GOP has been recently reformulated to utilize frame-level posterior probabilities directly from DNN-based acoustic models. The DNN-based GOP for the phoneme \( l_i \) is defined as:

\begin{equation}
\text{GOP}_{\text{DNN}}(l_i) = \frac{1}{t_2 - t_1} \sum_{t=t_1}^{t_2} \log \left( P(l_i \mid \mathbf{o}_t) \right)
\label{eq:gop_dnn}
\end{equation}

where \( P(l_i \mid \mathbf{o}_t) \) is the posterior probability of phoneme \( l_i \) given the acoustic observation \( \mathbf{o}_t \). This formulation eliminates the need for explicit likelihood computation, relying instead on frame-level outputs of the DNN model.

\subsection{Alignment-Free CTC Based GOP}

From \cite{cao2024framework}, we adapt the CTC-based alignment-free approach. This approach operates in two stages, taking speech features (\( \mathbf{O}_T \)) and the canonical transcription (\( L_{\text{canonical}} \)) as input. In this framework, the computation of GOP relies on the probability of the complete canonical transcription as well as that of the individual phonemes within the sequence.
The alignment-free GOP is formulated as:

\begin{equation}
\label{eq3}
\text{GOP}_{\text{alignment-free}} = \log \left( \frac{P(L_{\text{canonical}} \mid \mathbf{O}_T)}{P(L(i) \mid \mathbf{O}_T)} \right)
\end{equation}

This method works by first calculating the probability of the full canonical sequence \( P(L_{\text{canonical}} \mid \mathbf{O}_T) \). Then, we compute the probability of each phoneme in the canonical sequence by substituting it with other phonemes from the vocabulary or deleting the phoneme entirely to form a perturbed sequence. While this approach can deal with deletion and substitution errors in pronunciation assessment, it can impose a significant computational burden. This raises the question of whether it would be a good method at all. 

For a phoneme inventory of \(V = 39\) and the canonical transcription of "\textit{Would you like wine}" \textipa{/w U tS U l aI k w aI n/} with \(n = 10\) phonemes, computing the GOP score involves both substitution and deletion. Each phoneme can be substituted by \(V - 1\) alternatives or deleted once, leading to:
\[
\text{Total calculations} \sim n(V - 1) + n = 390
\]
This demonstrates how computational cost scales with transcript length and vocabulary size.

\subsection{Substitution-Aware Alignment-Free GOP}

In a Substitution-Aware CTC framework, GOP computation limits substitutions to a predefined set of phonetically confusable alternatives. For instance, if a learner is likely to pronounce /\textipa{\dh}/ as /d/, the alignment process allows /d/ as a valid alternative at the position originally labeled as /\textipa{\dh}/, rather than evaluating all possible substitutions. This is achieved using substitution mappings—sets of potential confusions—derived from phoneme clusters and common pronunciation errors made by non-native children learning a language.

Revisiting our example sentence, “\textit{Would you like wine},” if each phoneme is associated with at least three confusable phoneme pairs, the total number of calculations is reduced from 390 to 40, that is an approximate 90\% reduction in computation.

\subsubsection{Substitution Mapping Construction}
\label{sec:map}

A key aspect of our pronunciation assessment framework is the management of confusing phoneme pairs (or confusion sets) \cite{meng2009multilingual,oh2021hierarchical}.
The substitution mapping mechanism is a linguistically informed approach that limits potential phoneme replacements to acoustically or articulatorily similar pairs, preventing arbitrary substitutions. A handcrafted Phoneme Confusion Map is developed based on three main criteria:
(1) Phonetic Proximity, to capture natural articulatory relationships, stops, fricatives, and nasals are only substituted with phonemes that share similar places or manners of articulation (e.g., bilabial stops: /p/ → [/b/, /m/]); 
(2) common L2 learner errors, based on empirical observations of non-native speech patterns, reflecting frequent pronunciation mistakes (e.g., dental fricative substitutions: /\texttheta/ → [/ \textipa{\dh}/, /f/]);
and (3) phonological rules, prioritizing allophonic variants (e.g., the flap /\textfishhookr/ substituting for /t/ or /d/) and vowel mergers (e.g., /\textipa{I}/ → [/I/]), which account for common phonetic shifts across different learner populations \cite{cucchiarini2011error,cucchiarini2012my,kruitbosch2020pronunciation,wheelock2016phonological}. We applied this substitution mapping in two alignment-free GOP methods, which are described below.

\subsubsection{Phoneme-Adaptive Alignment-Free GOP (PA-AF GOP)}
In the numerator of Equation \ref{eq3}, the CTC loss is calculated by measuring how well the model's acoustic frames align with the original canonical phoneme sequence. This calculation involves the conventional forward-pass computation of $\alpha$-values (forward probabilities) to obtain the sequence likelihood. We denote this function as $\text{ctc\_loss}(p, y)$, where:
$p$ is a $(V \times T)$ matrix of per-frame posterior distributions, and $y$ is the ground-truth phoneme sequence of length $N$. 

In the denominator of Equation \ref{eq3}, we introduce a substitution-aware extension to the CTC forward pass, which computes CTC loss to evaluate the denominator term in GOP scoring while accounting for phonetically plausible mispronunciations. Unlike standard CTC loss, which estimates the likelihood of the reference phoneme sequence, this function incorporates two key adaptations. First, position-specific perturbations modify the target phoneme at a given position by either deleting it to model omission errors or substituting it with acoustically confusable phonemes from a predefined \textit{Phoneme Confusion Map}. Second, state-dependent token masking enforces substitution rules derived from linguistic knowledge by masking transitions to non-confusable phonemes during the dynamic programming computation of the forward-pass variable $\alpha$.

We define this modified function as $\text{ctc\_loss}(p, y, \text{pos}, M)$, where $\text{pos}$ denotes the index of the phoneme in $y$ that can be altered, and $M$ is a dictionary mapping each phoneme ID to its allowable substitutions. At the chosen index $\text{pos}$, the algorithm allows alignment to any phoneme in $M\bigl(y_{\text{pos}}\bigr)$, where $y_{\text{pos}}$ is the original phoneme at position $\text{pos}$. In the equation, a high GOP score indicates that substituting or deleting significantly decreases the log-likelihood, which means that the phoneme was correctly pronounced; a low GOP score indicates that substitutions or deletions have minimal impact on the log-likelihood, potentially indicating mispronunciation.

\subsubsection{Phoneme-Perturbed Alignment-Free GOP(PP-AF GOP)}

Unlike PA-AF GOP, which integrates substitution mechanisms directly within the CTC loss computation, PP-AF GOP handles phoneme substitutions and deletions externally by modifying the label sequences before computing the standard CTC loss. As in PA-AF GOP, we utilize the \textit{Phoneme Confusion Map} (Section~\ref{sec:map}) to guide these modifications.

The GOP score for each phoneme is computed based on the CTC loss difference between the original phoneme sequence and the perturbed phoneme sequence. For each phoneme, a set of perturbation sequences is created by (1) replacing the phoneme with mapped phonemes from the \textit{Phoneme Confusion Map}, generating a new phoneme sequence; and
(2) removing the phoneme from the sequence, creating an alternative sequence by omitting one phoneme. Finally, each perturbed sequence is evaluated based on the CTC loss of the acoustic model. The GOP score for each phoneme is computed as:

\begin{equation}
\label{eq:eq4}
    \text{GOP}(p) = \min(L_{\text{perturbed}}) - L_{\text{original}}
\end{equation}

\noindent 
where \( L_{\text{original}} \) is the CTC loss for the original phoneme sequence, and \( \min(L_{\text{perturbed}}) \) is the minimum CTC loss obtained across all perturbed phoneme sequences.

A higher GOP score (\(\geq 0\)) indicates that the original phoneme is more suited, while a negative GOP score suggests that a perturbation resulted in a lower CTC loss, meaning the phoneme is suboptimal or mispronounced. As in PA-AF GOP, we conducted our experiments using both Unrestricted Phoneme Substitutions (UPS) and Restricted Phoneme Substitutions (RPS) configurations.

\begin{figure}[ht!]
    \centering    \includegraphics[width=1.0\linewidth]{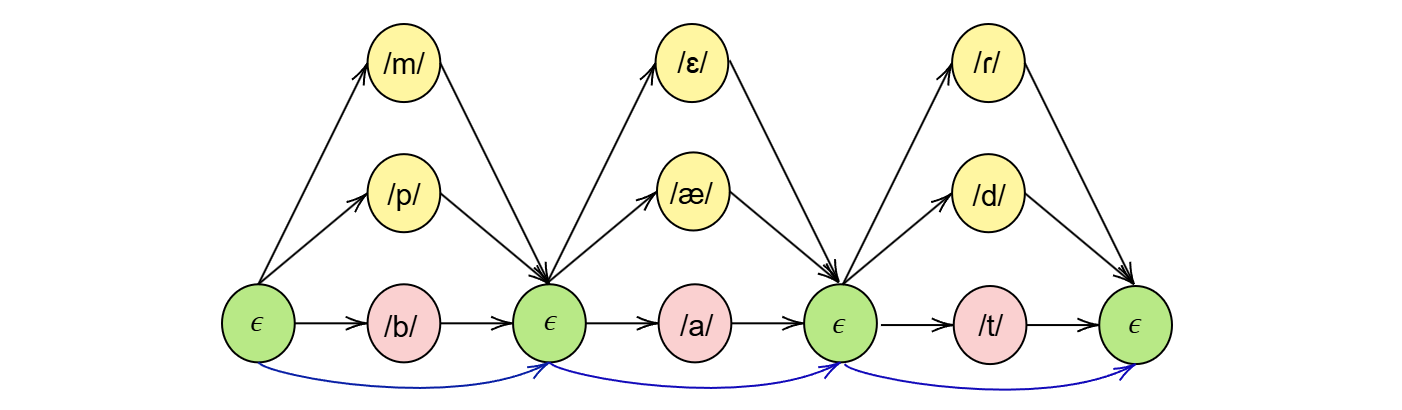}
        \vspace{-0.3cm}
    \caption{An illustrative example of the transition of the phonemes in the substitution mapping construction.}
    \label{fig: diagramalignmentfree}
    \vspace{-0.5cm}
        \end{figure}

An illustrative example of phoneme transitions in the substitution mapping is shown in Figure \ref{fig: diagramalignmentfree}. This diagram provides a conceptual understanding of how substitution mappings function within the alignment-free method. Each red target phoneme has possible substitutions (yellow), while deletions are indicated by blue lines. Given the canonical sequence \textipa{/b{\ae}t/}, possible sequences include \textipa{/p{\ae}t/}, \textipa{/m{\ae}t/}, and \textipa{/b{\ae}d/}, among others. Additionally, deletion-based variations, such as \textipa{/at/} (removal of \textipa{/b/}), \textipa{/bt/} (removal of \textipa{/a/}), and \textipa{/ba/} (removal of \textipa{/t/}), illustrate how phoneme deletion is handled in alignment-free.


\section{Experimental Procedure}

\subsection{Datasets}

\subsubsection{My Pronunciation Coach Dataset}

To answer our RQ, we conducted experiments with two datasets of L2 English speech. The first one, the MPC speech database \cite{cucchiarini2012my}, is particularly challenging as it contains L2 speech of children (124 in total) learning English in Dutch secondary schools. Child speech presents specific difficulties in terms of ASR over and above those related to L2 speech. Each recording in MPC includes 53 words and 53 sentences covering various English phonemes. Sessions are classified into four quality groups: Doubtful, Overloud, OK and Excellent. For this study, we selected OK and Excellent sessions (71 speakers: 38 males, 33 females), for a total of 3,130 utterances. As MPC lacks annotated mispronunciations, we introduced artificial errors by modifying phoneme sequences. These include replacing /\textipa{\dh}/ with /\textipa{d}/, /\textipa{T}/ with /\textipa{s}/, /\textipa{\ae}/ with /\textipa{e}/, /\textipa{\textturnv}/ with /\textipa{A}/, and simplifying diphthongs (e.g., /\textipa{eI}/ → /\textipa{e:}/, /\textipa{@U}/ → /\textipa{o}/).

\subsubsection{SpeechOcean762} 

To allow comparisons with previous research, we also used the SpeechOcean762 dataset \cite{zhang2021speechocean762}, an open-source corpus for pronunciation assessment that consists of 5,000 English utterances from 250 Mandarin-native speakers (125 adults, 125 children), with expert annotations at sentence, word and phoneme levels. Of 91,044 phoneme realizations, 3,401 were mispronunciations. We used all 5,000 utterances in our experiments.

\subsection{GOP Calculations}
We began our experiments by calculating GOP scores using the classical approach, as outlined in Equation~\ref{eq:gop_dnn}. To obtain forced alignment at the phoneme level, we employed a Kaldi-based Hidden Markov Model-Gaussian Mixture Model (HMM-GMM) system \cite{povey2011kaldi}, trained on the LibriSpeech \cite{7178964} 100-hour training dataset.

Regarding the forced alignment and alignment-free approaches, as described in Equations~\ref{eq3} and \ref{eq:eq4}, we generated pronunciation lexicons using representations of International Phonetic Alphabet (IPA) with the Phonemizer toolkit \cite{Bernard2021}. For the acoustic model, we utilized an openly available fine-tuned phoneme recognition model based on \cite{xu2021simple}, specifically \texttt{facebook/wav2vec2-xlsr-53-espeak-cv-ft}, hosted on HuggingFace. This multilingual model is built on the pretrained checkpoint \texttt{wav2vec2-large-xlsr-53} and has been fine-tuned on the CommonVoice dataset \cite{ardila2019common} to recognize phonetic labels across multiple languages.
The phoneme inventory of this model consists of 387 phonetic labels, excluding special tokens such as \texttt{<pad>}, \texttt{<unk>}, and sentence boundary markers (\texttt{<s>} and \texttt{</s>}).

We conducted experiments using both forced alignment and alignment-free methods, employing UPS and RPS configurations, to evaluate their effectiveness in mispronunciation detection. Experiments were performed with and without substitution mapping: UPS allows any phoneme to be replaced by another without constraints, while RPS incorporates substitution mapping to restrict phoneme replacements.\footnote{\tiny{\url{https://github.com/Aditya3107/GOP_MDD_Phonological.git}}}

\subsection{Evaluation Metrics}

We evaluated model performance using accuracy, precision, recall, F1-score, and Matthews Correlation Coefficient (MCC). Due to class imbalance in both MPC and SpeechOcean762, where correctly pronounced phonemes dominate, we optimized the threshold by selecting the GOP percentile that maximized MCC. Additionally, we reported the ROC AUC score at this threshold to assess classification effectiveness.

For SpeechOcean762, which includes human-annotated phoneme accuracy scores, we followed \cite{zhang2021speechocean762} and used second-order polynomial regression to model the relationship between GOP scores and human ratings. To evaluate predictive performance, we reported Pearson Correlation Coefficient (PCC) with confidence intervals and Mean Squared Error (MSE) for quantifying prediction accuracy.

\section{Results}

Table~\ref{tab:mpc-table} shows the experimental results on the MPC dataset. We report results for both forced alignment (FA) as the baseline, and two alignment-free approaches (PA-AF GOP, PP-AF GOP), with the latter evaluated under both RPS and UPS setups. Both PA-AF GOP and PP-AF GOP outperform FA GOP scores in most metrics, except recall, where FA achieves the highest value (0.929). However, FA has the lowest precision (0.165) due to the overclassification of correct pronunciations as mispronunciations, leading to a precision-recall trade-off.
The UPS setup, which considers all phonemes for substitution, generally results in higher recall and MCC values compared to RPS. MCC improves significantly, reaching 0.587 for PA-AF GOP and 0.595 for PP-AF GOP. However, the broader search space in UPS leads to a slight reduction in precision. The precision score for RPS in PP-AF GOP (0.509) is higher than in UPS and outperforms both PA-AF GOP setups, as well as FA. However, this comes with the cost of recall, which is significantly lower than in the UPS setup.




\begin{table}[!ht]
\caption{Performance evaluation with MPC}
\label{tab:mpc-table}
\vspace{-0.3cm}
\scriptsize
\setlength{\tabcolsep}{3pt} 
\begin{tabularx}{\columnwidth}{@{}l*{5}{>{\centering\arraybackslash}X}@{}}
\toprule
                       & \textbf{FA} & \multicolumn{4}{c}{\textbf{Alignment Free approaches}} \\
\cmidrule(lr){2-2} \cmidrule(lr){3-6}
                       & \textbf{Baseline}       & \multicolumn{2}{c}{\textbf{PA-AF GOP}}         & \multicolumn{2}{c}{\textbf{PP-AF GOP}} \\
\cmidrule(lr){3-4} \cmidrule(lr){5-6}
                       &                         & \textbf{RPS}     & \textbf{UPS}     & \textbf{RPS}    & \textbf{UPS}     \\
\midrule
AUC                    & 0.747                  & 0.869            & 0.941              & 0.883           & \textbf{0.949 }             \\
Accuracy               & 0.514                  & 0.888           & 0.898              & \textbf{0.902}          & 0.900              \\
Precision              & 0.165                  & 0.447           & 0.495              & \textbf{0.509}           & 0.500              \\
Recall                 & \textbf{0.929}                  & 0.511           & 0.822              & 0.581            & 0.831              \\
F1                     & 0.281                  & 0.477           & 0.618              & 0.543           & \textbf{0.624 }             \\
MCC                    & 0.242                  & 0.416           & 0.587              & 0.489           & \textbf{0.595  }            \\
AUC MCC\textsubscript{max}  & 0.698         & 0.720           & 0.864              & 0.759           & \textbf{0.869}              \\
\bottomrule
\end{tabularx}
\end{table}


Second, the experimental results of the SpeechOcean762 dataset are summarized in Table~\ref{tab:speechocean762}. Similar to the MPC dataset, FA achieves the highest recall (0.605). Among the alignment-free methods, only PP-AF GOP with UPS surpasses FA in key metrics such as accuracy, precision, F1, and MCC, achieving the best overall performance among all setups. Despite FA's strong AUC, it performs the worst in PCC compared to all alignment-free approaches. The highest PCC scores are achieved by the PP-AF GOP UPS setup (0.502 for high confidence and 0.488 for low confidence), followed by RPS (0.476 for high confidence and 0.461 for low confidence). In contrast, PA-AF GOP, both UPS and RPS, has lower PCC scores than PP-AF GOP in both setups but still outperforms the FA baseline. These results indicate that while FA retains an advantage in AUC at MCC$_{\max}$, alignment-free approaches—especially PP-AF GOP UPS—consistently achieve higher PCC scores and the lowest MSE (0.104), suggesting that they provide a more reliable phoneme-level pronunciation assessment by aligning better with human raters' phoneme scores.

\begin{table}[!ht]
\caption{Performance evaluation with SpeechOcean762}
\vspace{-0.3cm}
\label{tab:speechocean762}
\scriptsize
\setlength{\tabcolsep}{3pt} 
\begin{tabularx}{\columnwidth}{@{}l*{5}{>{\centering\arraybackslash}X}@{}}
\toprule
                       & \textbf{FA} & \multicolumn{4}{c}{\textbf{Alignment Free approaches}} \\
\cmidrule(lr){2-2} \cmidrule(lr){3-6}
                       & \textbf{Baseline}       & \multicolumn{2}{c}{\textbf{PA-AF GOP}}         & \multicolumn{2}{c}{\textbf{PP-AF GOP}} \\
\cmidrule(lr){3-4} \cmidrule(lr){5-6}
                       &                         & \textbf{RPS}     & \textbf{UPS}     & \textbf{RPS}    & \textbf{UPS}     \\
\midrule
AUC                    & 0.882                  & 0.853  & 0.859   & 0.887  & \textbf{0.916}   \\
Accuracy               & 0.924                 & 0.932  & 0.924   & 0.936 & \textbf{0.942}   \\
Precision              & 0.262                  & 0.239  & 0.261    & 0.273 & \textbf{0.322}   \\
Recall                 & \textbf{0.605}                  & 0.412  & 0.452   & 0.470 & 0.556   \\
F1                     & 0.366                  & 0.302  & 0.331   & 0.345 & \textbf{0.408}   \\
MCC                    & 0.365                  & 0.280  & 0.310    & 0.327 & \textbf{0.395}   \\
AUC MCC\textsubscript{max} & \textbf{0.771}             & 0.681  & 0.701   & 0.712 & 0.756    \\
\hline
PCC (low conf)         & 0.279                  & 0.404  & 0.437    & 0.461 & \textbf{0.488}   \\
PCC (high conf)        & 0.297                  & 0.419  & 0.424   & 0.476 & \textbf{0.502}   \\
MSE                    & 0.125                  & 0.113  & 0.112    & 0.107 & \textbf{0.104}   \\
\bottomrule
\end{tabularx}
\end{table}


While previous work on similar methods reports a PCC of 0.56 \cite{cao2024framework}, those approaches involve building a complete MDD model using multidimensional GOP scores. In contrast, our method is optimized for real-time pronunciation error detection, emphasizing efficiency and simplicity. To the best of our knowledge, the highest PCC reported for phoneme-level performance on SpeechOcean762 is 0.69 \cite{chao2023hierarchical}. However, this methodology cannot be directly compared to ours, as it includes additional contextual information as input and employs a more advanced multi-task training framework.

\section{Discussion and Conclusion}


The results obtained in this work answered our RQ, demonstrating that phoneme clustering and learner-specific error modeling can reduce computational costs in an alignment-free CTC-based MDD system. However, it is important to note that these approaches may also lead to a decline in performance compared to unrestricted phoneme substitutions.
This decline occurs because phoneme recognition models are not optimized for predefined phoneme clusters, limiting their ability to generalize pronunciation variations. 
Our results (Table \ref{tab:mpc-table} and Table \ref{tab:speechocean762}) confirm that Substitution-Aware Alignment-Free GOP methods can provide a balance between efficiency and accuracy, making them more suitable for use in CAPT applications for users, in which instantaneous feedback is needed.


We observed that PA-AF GOP is more affected by RPS compared to PP-AF GOP. A possible explanation is that PA-AF GOP modifies the CTC forward pass internally, meaning phoneme substitutions are constrained within the CTC alignment process. When using RPS, if the correct mispronunciation is not in the predefined confusing phoneme pairs, the CTC function cannot align it correctly, leading to higher errors. This also suggests that PP-AF GOP is more robust to RPS.

We also observed that the FA baseline has high recall values in both datasets. A potential reason for this could be errors in forced alignment due to acoustic variability in non-native speech in both of our datasets. Minor acoustic variations can lead to over-detection of mispronunciations, where correctly pronounced phonemes are misclassified as errors. This results in a high recall but low precision.

One important aspect of our work is that we considered common substitution and deletion errors made by non-native speakers, but no insertion errors. This is because, unlike substitutions and deletions, insertions are often unpredictable and may include non-lexical sounds such as "umm" or "hmm". Furthermore, insertions are generally less frequent in read-aloud tasks with short utterances, as in our study, than in spontaneous conversations. 

The methods that we proposed in this work can also be easily applied to other languages. Expanding the phoneme search space could enhance performance and flexibility in language learning applications. Future research could refine phoneme selection strategies to balance accuracy and efficiency. Additionally, our model-agnostic approach allows integration into any CTC-based phoneme recognition system.



\section{Acknowledgements}
This publication is part of the project Responsible AI for Voice Diagnostics (RAIVD) with file number NGF.1607.22.013 of the research programme NGF AiNed Fellowship Grants which is financed by the Dutch Research Council (NWO).

\bibliographystyle{IEEEtran}
\bibliography{mybib}

\end{document}